\documentclass[a4paper]{ieee}
\usepackage{times}
\usepackage{graphicx}
\usepackage{algorithm}
\usepackage{algorithmic}
\usepackage{epsfig}
\usepackage{amsmath,amssymb}
\usepackage{balance}

\def\argmax{\operatornamewithlimits{arg\,max}}

\begin{document}

\title{A Combined LIFO-Priority Scheme for Overload Control of
  E-commerce Web Servers} 

\author{Naresh Singhmar ~~~ Vipul Mathur ~~~ Varsha Apte ~~~
  D.~Manjunath\\
  Indian Institute of Technology - Bombay\\
  Powai, Mumbai, 400 076, India\\
  \texttt{\{naresh,vipul,varsha\}@cse.iitb.ac.in, dmanju@ee.iitb.ac.in }
  }

\maketitle

\begin{abstract}
  E-commerce Web-servers often face overload conditions during which
  revenue-generating requests may be dropped or abandoned due to an
  increase in the browsing requests. In this paper we present a
  simple, yet effective, mechanism for overload control of E-commerce
  Web-servers. We develop an E-commerce workload model that separates
  the browsing requests from revenue-generating transaction requests.
  During overload, we apply LIFO discipline in the browsing queues and
  use a dynamic priority model to service them. The transaction queues
  are given absolute priority over the browsing queues. This is called
  the \emph{LIFO-Pri} scheduling discipline. Experimental results show
  that LIFO-Pri dramatically improves the overall Web-server
  throughput while also increasing the completion rate of
  revenue-generating requests. The Web-server was able to operate at
  nearly 60\% of its maximum capacity even when offered load was 1.5
  times its capacity. Further, when compared to a single queue FIFO
  system, there was a seven-fold increase in the number of completed
  revenue-generating requests during overload.
\end{abstract}


\section{Introduction}
\label{sec:Intro}
The capacity of a Web-server is measured in terms of the rate of
requests/second that it can fulfill. When the request rate to a
Web-server exceeds its capacity, the server is overloaded, its
response time increases to an unacceptable level, and requests start
timing out, i.e., they are abandoned, typically after some service has
been received. Abandonments lead to retries, and the effective load on
the server increases further. In this situation, in the absence of an
overload control mechanism, the server ends up being busy doing
unproductive work and the throughput degrades.  E-commerce
Web-servers, e.g., retail Web sites, often experience such overload
situations, triggered by events such as closing time of a sale or
intense shopping days \cite{Christmas}.

Occurrence of overload situations can be minimized by appropriately
sizing the server centers and by using techniques such as load
balancing. However, overloads are not completely
avoidable---unexpected consumer demand, partial server failures, or
other such events can trigger unexpected overloads. We therefore need
mechanisms to protect the Web-server from being pushed to an
unproductive state during overloads. In this paper we propose and
experimentally analyze one such mechanism. Specifically, we focus on
E-commerce Web-servers, e.g., the server for an on-line store.  For
such Web-servers, the requirement is not only to be productive during
overload, but to be able to differentiate between direct
\emph{revenue-generating requests} and \emph{browsing requests} that
generate revenue only indirectly. On typical shopping Web sites, the
load due to the browsing requests far exceeds that of the
revenue-generating requests and it is imperative that the browsing
requests do not prevent revenue-generating requests from getting
completed.

Overload control of telecommunication switches has been studied
extensively, e.g., \cite{Doshi86}, and some of the principles
developed there can be applied to Web-servers. However, there is an
important difference between a Web-server and a telecommunication
switch. The former is typically modeled as a single queue (with single
or possibly multiple servers) while the latter is a multi-queue
system.  Furthermore, since the servers take the form of processor
threads, the service rate of the servers is a decreasing function of
the number of active servers. Thus it is not clear if the overload
control methods developed for telecommunication switches will be
directly applicable.  Therefore experimental evaluation like the one
that we do in this paper is necessary.

Overload control of Web-servers has gained much attention in the
recent past.  Approaches include admission control \cite{Cherkasova02}
or sophisticated scheduling policies \cite{Chen02}, or both
\cite{Elnikety04}. The fact that Web usage is session-oriented has
been recognized, and several overload control mechanisms are based on
that. A mechanism that does not admit new sessions at overload was
proposed in \cite{Cherkasova02}. The mechanism proposed by
\cite{Chen02} employs a dynamic weighted fair sharing policy to
process requests from those sessions that are more likely to complete.
This is done by dynamically adjusting the weights of the queues, as
calculated by maximizing a productivity function. Elnikey et
al~\cite{Elnikety04} propose and implement an admission control and
request scheduling policy, in which the resource requirement of a
request is estimated by an external entity, and admission control is
done based on that. Furthermore, a shortest job first scheduling
approach is utilized for improving response times. A control theory
based approach to overload control is described in
\cite{Abdelzaher02}. The authors use a feedback control loop based
mechanism to prevent overload by monitoring the utilization of server
resources and switching to a degraded QoS level in overload
conditions. However, their solution is meant primarily for static
content as it relies on the availability of an alternate `degraded'
set of objects to be served. Thus, it does not take into account the
variable execution time of scripts that are involved in serving
dynamic content. Hence the approach of \cite{Abdelzaher02} is not
directly applicable to an E-commerce scenario such as the one we have
considered here.

Our survey suggests that although a number of mechanisms have been
proposed, none of the work focuses on the essential difference between
revenue-generating and browsing requests, that are a characteristic of
an E-commerce Web site.  In our work, we specifically recognize this
difference, and work from there. We assume that the ultimate goal of
an E-commerce Web site is to complete as many revenue-generating
requests as possible---any work that an E-commerce Web-server does
should be in support of this final goal. We propose a simple
combination of priority queuing and last-in-first-out (LIFO)
scheduling during overload, to achieve this goal. We have implemented
and analyzed our mechanism experimentally, by emulating a typical
E-commerce Web site.  We use a session-based workload model that
emulates realistic user behavior---variable abandonments, variable
retries, and session abandonments as a result of request abandonments.
We show that our mechanism performs well under all such realistic
conditions.

Note that the use of LIFO for overload control when dealing with
impatient customers is not new and has been proposed for
telecommunication systems. Doshi and Heffes~\cite{Doshi86} provide an
excellent analysis of this family of service disciplines for overload
control.  They have analytically shown that LIFO based schemes are
more attractive at overload from both throughput and delay points of
view.  Note though, that in the absence of overload, the response time
of LIFO will have a higher variance than that of FIFO and can hence
cause more abandonments than FIFO. In fact, we have experimental
results to show that this does happen in the case of Web-servers.

The rest of the paper is organized as follows.  In
Section~\ref{sec:Proposal} we propose an E-commerce workload model and
describe our LIFO-priority based overload control mechanism.  In
Section~\ref{sec:Experiment} we describe the experimental setup and
discuss the results. We conclude in Section~\ref{sec:Conclude} with
discussions and suggestions for future work.


\section{Proposed Overload Control Scheme for E-Commerce Web-Servers}
\label{sec:Proposal}
The goal of an E-commerce Web site, is revenue generation, which it
achieves by allowing visitors to browse through its merchandise (if it
is a retailing Web site), and then buy. Since a large fraction of the
browsing visitors do not intend to buy, it is important that those
that have shown the intent to buy by beginning the buying process must
be helped to complete the transaction without timing out and
abandoning the transaction. This is especially important during
overload conditions when the server strains under an increased
overhead. Before describing the overload control scheme we present our
model about a typical E-commerce workload.

\subsection{An E-Commerce Workload Model}
\label{E-Com-Workload}

\begin{figure} 
  \begin{center}
    \includegraphics[width=3.2in]{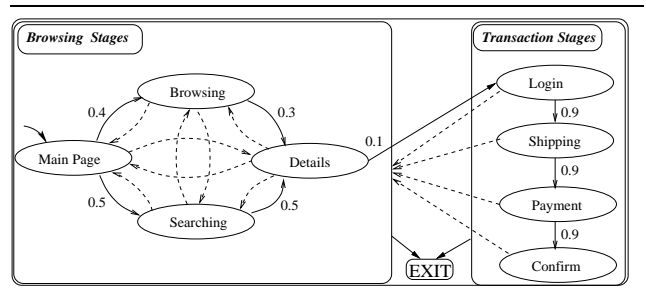}
    \caption{FSM diagram of a session representing a retail Web site.
      We also show example transition probabilities for the case of
      the FSM representing a Markov Chain. The self-loops and the exit
      probabilities from each stage are not shown. In our experiments
      the transitions shown as dashed lines are assumed to have zero
      probability.}
    \label{bro_buy_stages}
  \end{center}
\end{figure}

We assume that in an E-commerce Web site most of the users browse the
site for some time and leave, while a few of these browsing users
proceed towards a revenue-generating transaction that is a multi-step
(multiple Web page) process.  For example, in an online retail site,
the user first visits the home page and then possibly browses or
searches through the catalog. If the preferred product is available
then more details about that product may be sought. We term these
requests as the \emph{browsing requests}. Most of the users leave the
site at this point; few who have the intention of buying some product
proceed to the first step in a sequence of transactions, e.g., the
`login' page. From this point onward, the user is led through a
multi-stage sequenced transaction (involving, e.g., entering payment
and shipping details), usually culminating in a `confirm' request,
that finalizes the transaction. We term these the \emph{transaction
  requests}. This movement of the user between the different types of
pages can be represented by a finite state machine as shown in
Fig.~\ref{bro_buy_stages}. To construct a tractable model that can
simplify simulation and analysis, we assume that the transitions
between the states are memoryless and that the probabilities are
stationary. Thus the user behavior can be modeled as a stationary
finite state Markov chain with states corresponding to the pages.

\subsection{LIFO-Pri Overload Control Algorithm}
\label{sec:LIFO-Pri-describe}
Recall that our objectives are twofold---(1) maximize the throughput
of revenue-generating requests while (2) improving overall throughput
of the Web-server during overload. The mechanism that we propose in
the sequel will be called \emph{LIFO-Pri}.

To achieve the first goal of maximizing the throughput of
revenue-generating requests, we employ a priority mechanism. Separate
queues are maintained for each type of request. The transaction
request queues are given a simple non-preemptive priority over the
browsing request queues.  We make a simplifying assumption that we
would never want to serve any browsing request if a transaction
request is waiting to be served. Between the transaction queues, the
queue for the last request, e.g. `confirm', in the multi-step
transactions has the highest priority. The queue for the request, e.g.
`payment', just before `confirm' has the second highest priority, and
so on.

To achieve the second goal of maximizing the overall throughput, we
propose a load-based LIFO mechanism---a FIFO policy during normal load
and a LIFO policy during overload. As noted earlier LIFO based
policies provide better throughput and delay performance at overload
as compared to FIFO.  This can be explained as follows. Since the mean
delay at overload is high, the high variance of the delay works in our
favor by having more requests that do not time out than would happen
with FIFO.

We make the reasonable assumption that overload is primarily due to
browsing requests. Hence we employ LIFO during overload only for the
browsing queues while serving the transaction queues according to
FIFO.

We also propose a dynamic priority mechanism for selecting requests
from the browsing queues to allow those that may have a higher chance
of leading to a transaction request to complete with a higher
probability.  We use dynamic priorities because static absolute
priorities can lead to starvation of low priority queues. The proposed
scheme is as follows. For the browsing queues, two different
attributes are maintained for each queue:
\begin{itemize}
\item Number of pending requests in that queue ($N_{i}$).
\item Utility of that queue ($U_{i}$).
\end{itemize}

The queue priority at any time is then given by $U_i \times N_i$. The
\emph{utility} is an indicator of the relative importance of the
queues. This {utility} could, for example, be based on the `revenue
generation potential', i.e., if the `details' page request is more
likely to lead to a buy request than a `search' page request, then the
`details' page can be given a higher {utility}. The values for the
utility may be obtained from the Markov chain describing the user
behavior. By including the queue length in obtaining the priority, we
prevent the lower priority queues from getting starved.

\begin{algorithm}
  \caption{\emph{LIFO-Pri}}
  \label{algo:LIFO-Pri}
  SET\_DISCIPLINE:
  \begin{algorithmic}
    \WHILE {alive}
      \STATE CPU\_Util $\Leftarrow$ Utilization measured over an interval
      \IF{(CPU\_Util $>$ CPU\_Upper\_Threshold) \textbf{AND} (Browsing\_Policy $=$ FIFO)}
        \STATE Browsing\_Policy $\Leftarrow$ LIFO
      \ENDIF
      \IF{(CPU\_Util $<$ CPU\_Lower\_Threshold) \textbf{AND} (Browsing\_Policy $=$ LIFO)}
        \STATE Browsing\_Policy $\Leftarrow$ FIFO
      \ENDIF 
    \ENDWHILE
  \end{algorithmic}

  DYNAMIC\_PRIORITY:
  \begin{algorithmic}
    \WHILE {alive}
      \IF{(A worker thread is available) \textbf{AND} (At least one queue has a pending request)}
	\FORALL{$1 \leq i \leq $ Number of queues}
	  \STATE $DP_i \Leftarrow N_i \times U_i$
	\ENDFOR
	\STATE $Q \Leftarrow \argmax_i (DP_i)$
	\STATE Read a request from queue $Q$ according to current service discipline.
	\STATE Assign worker thread to request.
      \ENDIF
    \ENDWHILE
  \end{algorithmic}
\end{algorithm}

The service discipline used by {LIFO-Pri} for the browsing requests
depends on the CPU utilization. If the CPU utilization crosses a
predefined \emph{upper threshold}, then it starts serving the browsing
requests according to LIFO, and it continues with this discipline
while the CPU utilization is above \emph{lower threshold}.  Recall
that the transaction requests are always served in FIFO order.

The above discussion is summarized in Algorithm~\ref{algo:LIFO-Pri}.
Note that the two parts of the algorithm---SET\_DISCIPLINE and
DYNAMIC\_PRIORITY have to be executed in parallel, typically by
separate threads.


\section{Experimental Results and Discussions}
\label{sec:Experiment}
\begin{figure} 
  \begin{center}
    \includegraphics[width=3.2in]{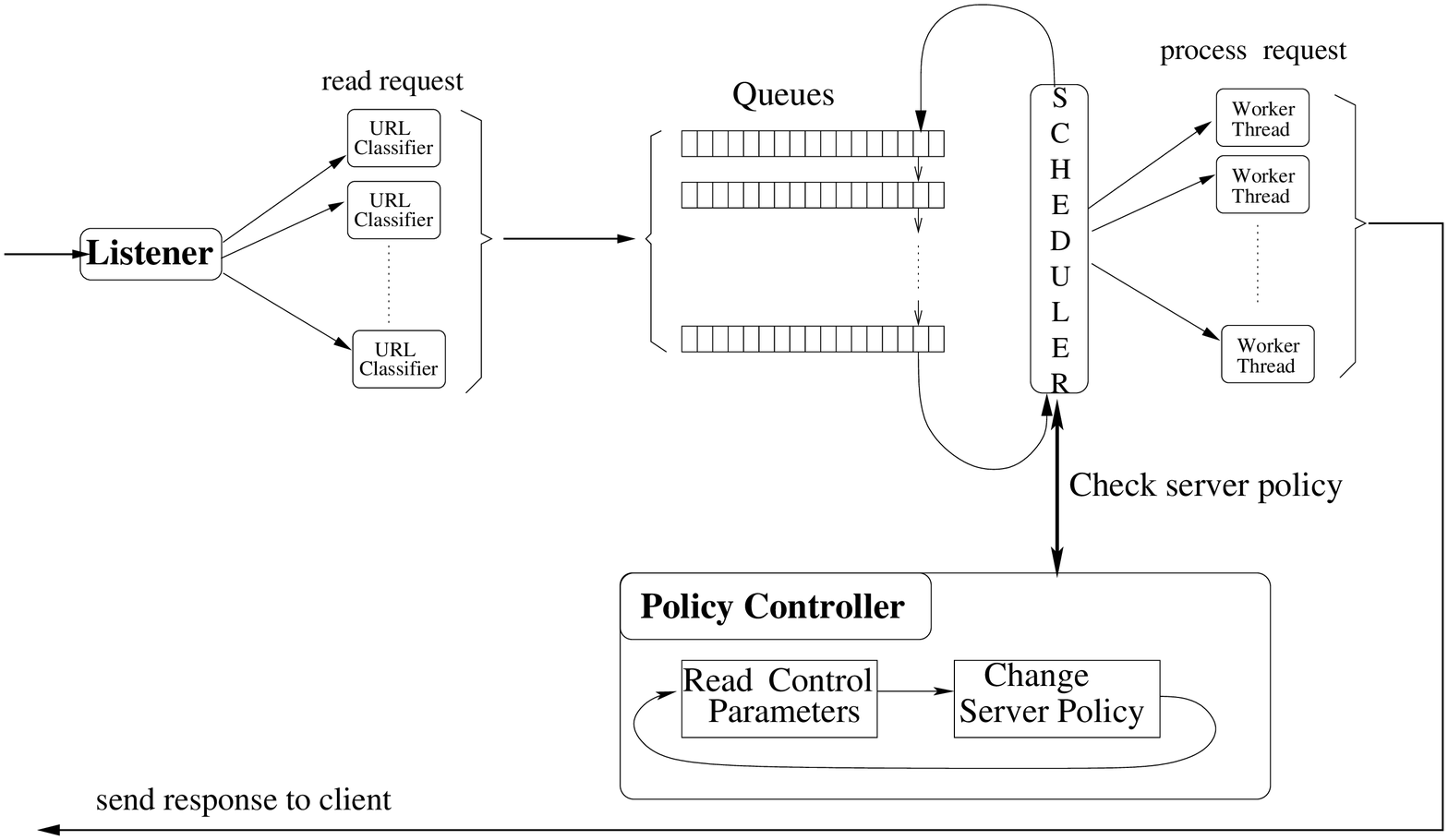}
    \caption{Web-server architecture}
    \label{Web_server}
  \end{center}
\end{figure}

The overload control policy as described in the above section was
implemented in a Web-server. The Web-server architecture is as
depicted in Fig.~\ref{Web_server}.  Experiments were carried out to
verify the performance of our overload control mechanism, by varying
load on the Web-server that we have built. The experiments done can be
divided into two parts:
\begin{itemize}
\item Experiments to compare FIFO and LIFO service order.
\item Experiments with an E-commerce setup to test the LIFO-Pri
  policy.
\end{itemize}

The first set of experiments separately characterize performance of
LIFO and FIFO under non-overload and overload conditions on the
Web-servers. These experiments offer several insights that will be
discussed later in this section.  The second set of experiments test
the effectiveness of the LIFO-Pri overload control mechanism.

The test-bed contains a server and a client machine. The server machine
is based on an Intel P-IV 1.6~GHz CPU with 256~MB~RAM, running Debian
GNU/Linux Sid. The Web-server runs on this machine with a maximum
limit of 30 worker threads.  It must be noted that the priority
assignment is only for the assignment of a worker thread and the
transaction queues are not given priority in execution by the
operating system. The client machine is based on an Intel
P-IV~2.4~GHz CPU with 256~MB~RAM running Debian GNU/Linux Sid. The
client is used to generate load on the Web-server using
\texttt{httperf} \cite{httperf98}.

        
\subsection{Comparison of FIFO and LIFO }
\label{sec:comparison}

Since we are specifically comparing the performance of LIFO and FIFO
service disciplines, we carried out a set of experiments on a basic
Web-server with a single queue and not with the E-commerce setup model
of Fig.~\ref{bro_buy_stages}. The load is generated by repeatedly
making a request for a \emph{CPU-intensive} CGI file. The distribution
of the inter-arrival time between requests is exponential.

\subsubsection{Experimental Setup}
\label{sec:setup1}
The Web-server is configured with a single queue with a buffer
capacity of fifty. To compare the FIFO and LIFO approaches, we repeat
the experiments with the following three different service policies.
In the first case, called \emph{Always-FIFO}, the Web-server always
serves the requests in FIFO order. In the second case, termed
{Always-LIFO}, the Web-server always serves the requests in LIFO
order. In the third case, that we call \emph{LIFO-at-overload}, the
service discipline alternates between LIFO and FIFO as is done in
LIFO-Pri.


\subsubsection{Results}
\label{sec:results}

The experiments were performed to study the server response as a
function of increasing load. In this set of experiments, we use a fixed
timeout value for all the requests.

\begin{figure} 
  \begin{center}
    \includegraphics[width=2.2in,angle=270]{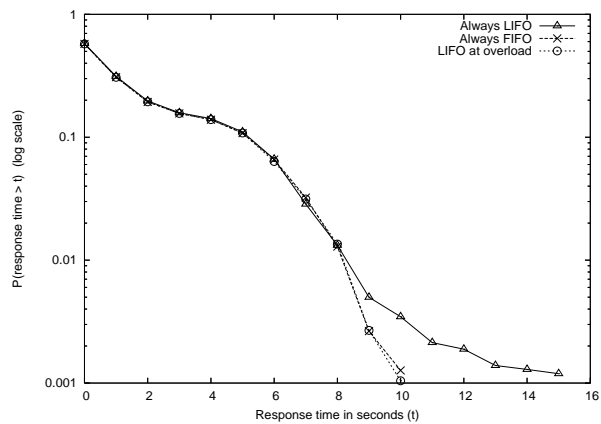}
    \caption{Response time distribution at $\rho=0.941$ with a
      timeout of 20~seconds.}
    \label{Log_LF_rate2-8_TO18_dis}
  \end{center}
\end{figure}

\begin{table*}
  \begin{center}
    \vspace*{0.1in}
    \begin {tabular}{|l|c|c|c|} \hline
      \multicolumn{4}{|c|}{Timeout of 40~seconds} \\ \hline \hline
      Percentage (\%) & Always-FIFO & Always-LIFO & LIFO-at-overload
      \\ \hline  
      Requests Completed & 86.7 & 84.4 & 84.6 \\
      Requests Timeout   &  0.0 &  2.3 &  2.0 \\
      Requests Dropped   & 13.3 & 13.4 & 13.4 \\ \hline \hline 
      \multicolumn{4}{|c|}{Timeout of 20~seconds} \\ \hline \hline 
      Requests Completed & 21.9 & 81.0 & 76.8 \\
      Requests Timeout   & 64.9 &  5.4 &  9.7 \\
      Requests Dropped   & 13.3 & 13.6 & 13.4 \\ \hline
    \end{tabular}    
    \caption{Comparison of FIFO and LIFO based service disciplines
      in a single queue system. Server throughput at $\rho=1.47$ with 
      a timeout of 40 and 20~seconds. }
    \label{LOG_lifo_fifo}
  \end{center}
\end{table*}

Denote the server intensity (ratio of arrival rate to service rate) by
$\rho$. When the offered load is below the capacity of the server,
i.e., $\rho < 1.0$, the number of requests that are either dropped or
timed out is almost zero for all the three cases.
Fig.~\ref{Log_LF_rate2-8_TO18_dis} shows the unconditional
complementary distribution of the response time\footnote{All
  response time distribution graphs in this paper are the unconditional
  complementary distributions. This allows us to treat the response
  time of the timed out or dropped requests to be infinity.}  for
$\rho=0.941$.  Observe the longer tail for the case of {Always-LIFO}
implying that a significant fraction of requests have a long response
time.  This effect is not seen when {LIFO-at-overload} is used.
Thus, using LIFO is not appropriate when the load is less then the
capacity of the Web-server.


When the offered load is higher than the capacity of the server,
requests are dropped or are abandoned due to timeouts. We consider two
timeout values---40~seconds and 20~seconds to model less patient
customers. It can be seen in Table~\ref{LOG_lifo_fifo} that the
percentage of requests dropped is almost identical for all the three
service schemes but the abandonment rate depends significantly on the
timeout value.  First, consider the case when the timeout value is 
40~seconds. Here, as is to be expected, {Always-FIFO} has the lowest
percentage of abandoned requests. A large timeout value favors FIFO,
because the FIFO response time does not have the ``long tail'' of
LIFO. Note that even for the same average queue length, LIFO may
result in much larger response time values than FIFO (a request that
gets ``pushed'' to the end of the queue may never get served, and will
eventually timeout). With a 20~second timeout, the {Always-FIFO}
policy now shows a much larger abandonment rate than the LIFO
policies. Further, the FIFO policy is able to achieve only 21.9\%
success rate as opposed to about 80\% for the LIFO policies.

\begin{figure*}
  \begin{center}
    \begin{tabular}{cc}
      \parbox{3.2in}{
        \centering \
        \psfig{figure=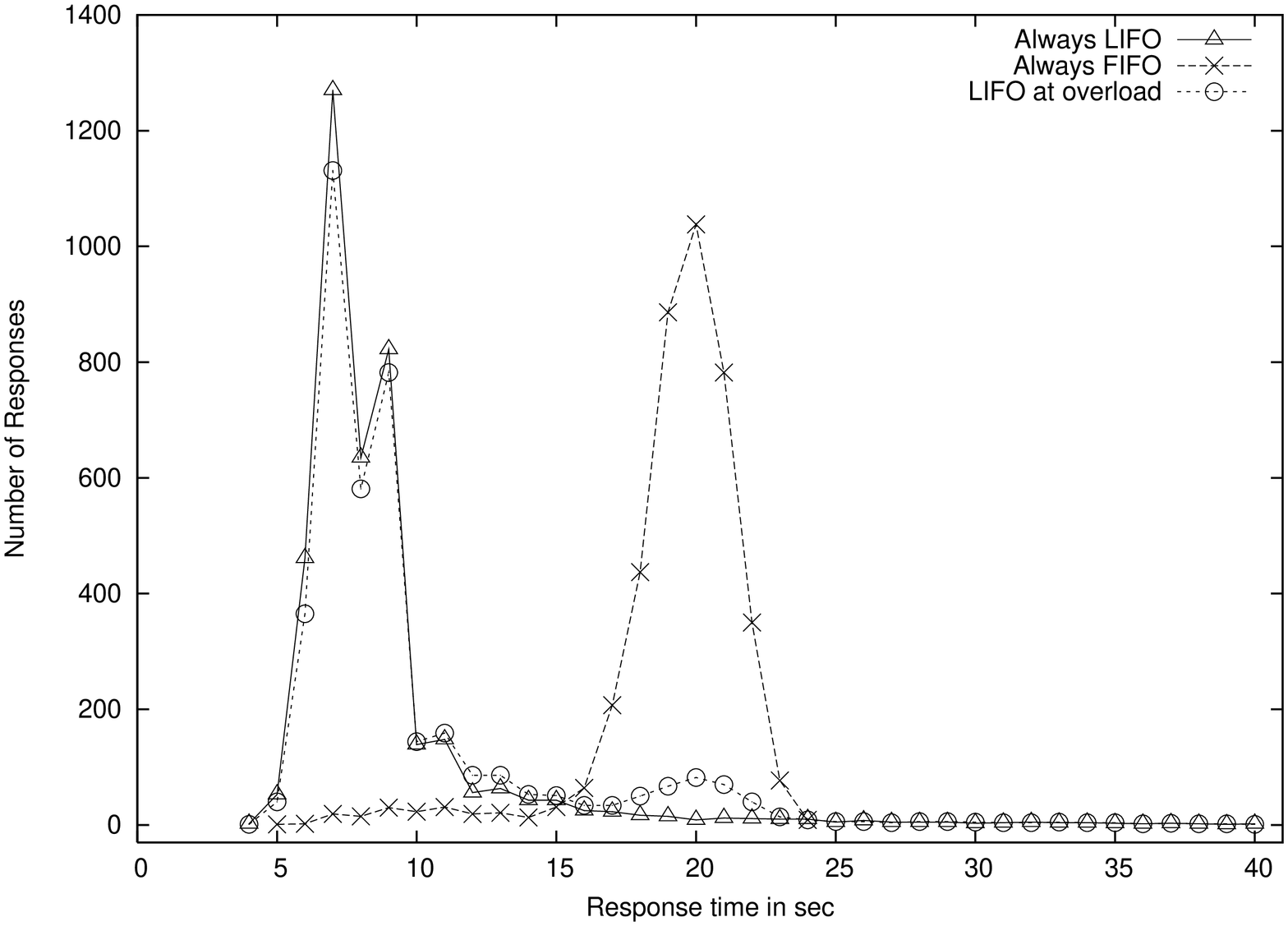,height=3.2in,angle=270}
        }  &
      \parbox{3.2in}{
        \centering \
        \psfig{figure=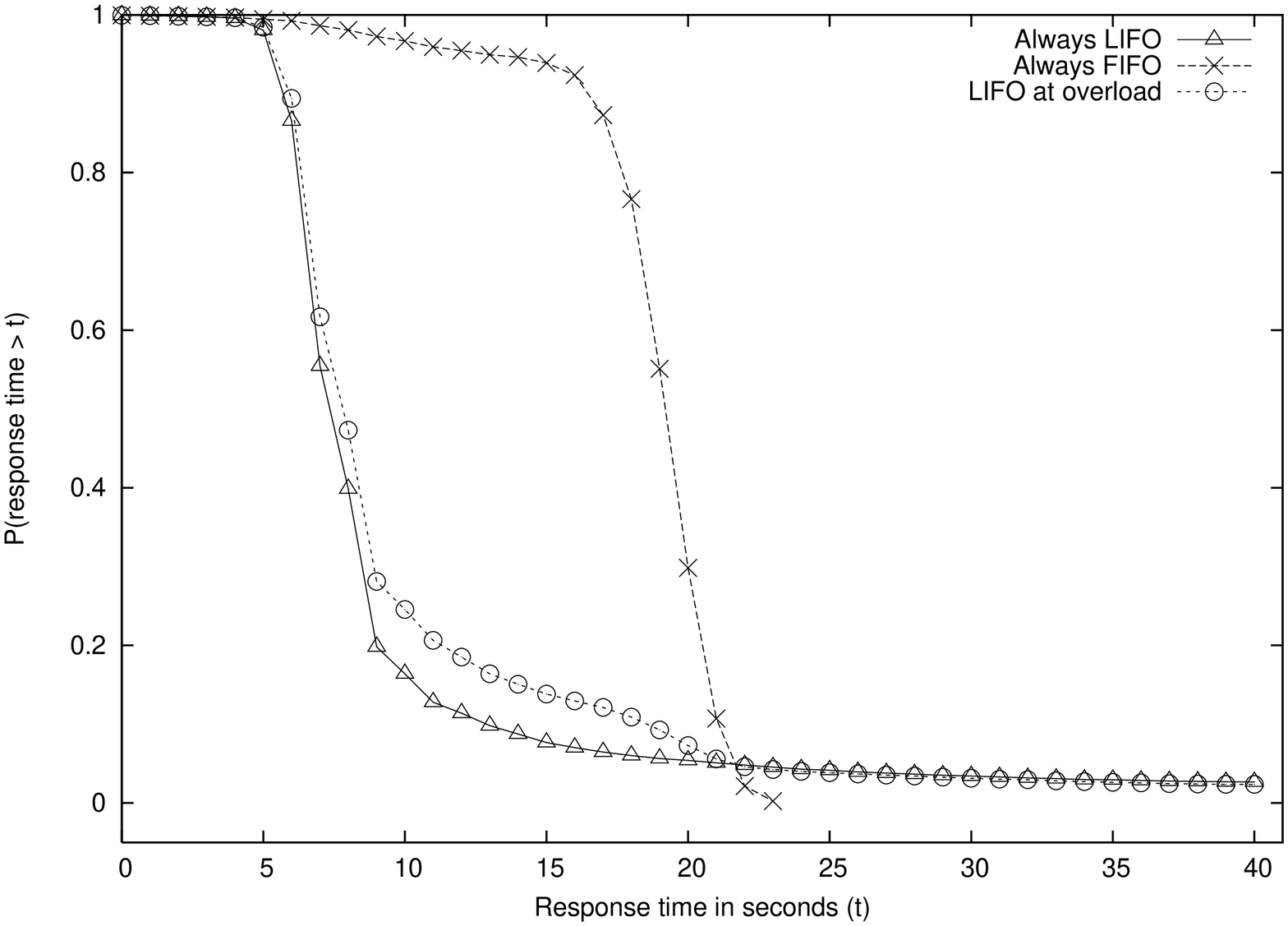,height=3.2in,angle=270}
        } \\
      \multicolumn{2}{c}{} \\
      Histogram & Distribution
    \end{tabular}
  \end{center}
  \caption{Response time histogram and distribution for $\rho=1.47$
    and a timeout of 40~seconds.}
  \label{plot:LOG_LF_rate5_TO40}  
\end{figure*}

\begin{figure*}
  \begin{center}
    \begin{tabular}{cc}
      \parbox{3.2in}{
        \centering \
        \psfig{figure=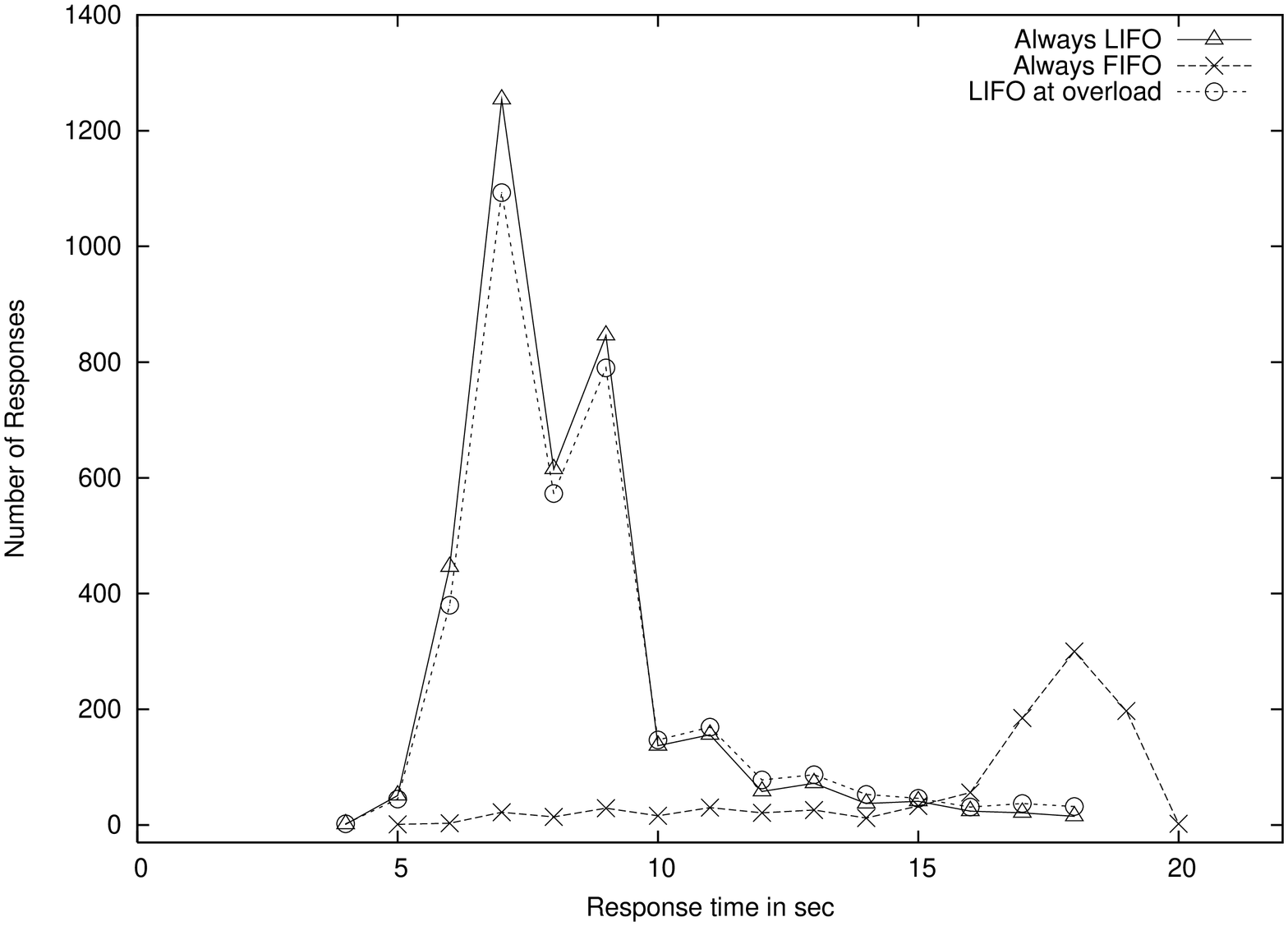,height=3.2in,angle=270}
        }  &
      \parbox{3.2in}{
        \centering \
        \psfig{figure=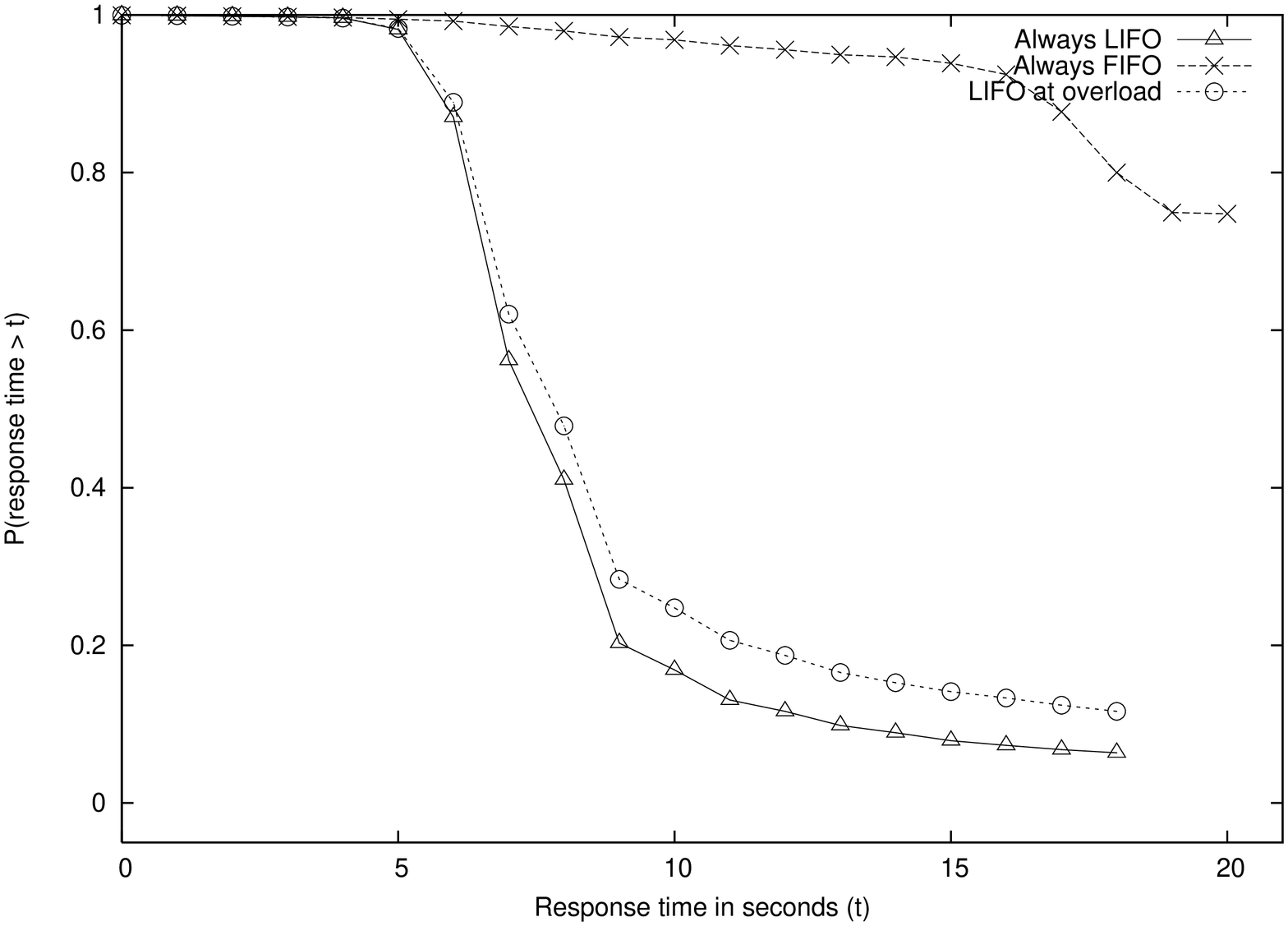,height=3.2in,angle=270}
        } \\
      \multicolumn{2}{c}{} \\
      Histogram & Distribution
    \end{tabular}
  \end{center}
  \caption{Response time histogram and distribution for $\rho=1.47$
    and a timeout of 20~seconds.}
  \label{plot:LOG_LF_rate5_TO20}  
\end{figure*}

Fig.~\ref{plot:LOG_LF_rate5_TO40} shows the response time histogram
and distribution for the case of $\rho=1.47$ and a timeout of 40~seconds.
Observe that for {Always-FIFO} the mode is at 20~seconds.
Also see that for {Always-FIFO} all the requests have a response time
of less than 24~seconds (which explains no abandonments, since the
timeout is 40~seconds). For the two LIFO-based policies we observe two
interesting phenomena---the mode occurs at about 7~seconds but a
significant number of the requests have a very large response time,
even as large as 40~seconds. This is also reflected in the long tail
of the LIFO response time distribution.

Fig.~\ref{plot:LOG_LF_rate5_TO20} shows the histogram and the
distribution of the response time with a timeout of 20~seconds.
Comparing with the 40~second timeout case, we observe that the
difference in histograms for {Always-LIFO} and for {LIFO-at-overload}
does not change significantly with the timeout value except that the
tail is shortened. However, for the case of {Always-FIFO} the mode of
the distribution is at about 18~seconds. Also, for {Always-LIFO} and
for {LIFO-at-overload}, nearly 80\% of the requests have a response
time of less then 10~seconds, whereas for {Always-FIFO}, less than 5\%
of requests experience this response time.

Thus by using {LIFO-at-overload} approach we have achieved not only
higher throughput, but also significantly better response time
distribution at higher load.


\subsection{Experimental Analysis of LIFO-Pri}
\label{sec:LIFO-Pri-Analysis}
In the experiments described in the previous section the workload
consisted of a random sequence of requests for URLs and did not
correspond to a transaction. We verified the claim that using {LIFO}
service discipline improves the performance of a Web-server during
overload in the presence of impatient users. We now present results of
experiments that were performed to test the proposed LIFO-Pri
mechanism.

\subsubsection{Experimental Setup}
For validating our mechanism, we set up a Web site that emulates the
characteristics of a typical E-commerce Web site as per our model of
Fig.~\ref{bro_buy_stages}. Some of the possible transitions in the
model were assigned a probability of zero so as to minimize the effect
of `unknown' factors in the controlled experiments.

The eight types of pages shown in Fig.~\ref{bro_buy_stages} are
generated using Perl CGI scripts that have interleaved random
busy and waiting periods. The busy periods represent local processing
and the waiting periods represent time spent in the back-end server
calls such as database lookups. Table~\ref{cgi-exe-time} shows the
mean execution times (including the delay in servicing back-end
requests) of these CGI scripts.

We use \texttt{httperf} with the \texttt{--wsesslog} option to
generate the E-commerce workload.  i.e., \texttt{httperf} reads
session descriptions from a file of 1000 randomly generated session
descriptions according to the Markov chain shown in
Fig.~\ref{bro_buy_stages} and keeps cycling through them until a
specified total number of sessions have been completed.  This is
necessary because we did not have a load generator that could generate
such a randomly distributed workload. Each session consists of a
sequential set of requests which must be completed for the session to
succeed. The session arrival process is modeled to be Poisson.

As in the previous section, we model the `impatience' of the users by
using timeouts for the requests.  \texttt{httperf} supports two kinds
of timeouts. The basic timeout is called \texttt{--timeout} and it is
the amount of time that the load generator waits for a server
reaction, i.e., forward progress must be made within this timeout
value while creating a TCP connection, sending a request, waiting for
or receiving a reply.  An additional \texttt{--think-timeout} is added
to the basic timeout while waiting for a reply after issuing a
request. This is used to allow for the additional response time that
the server might need to \emph{initiate} sending a reply for a
request, since we are running time-consuming CGI scripts and not
merely fetching a static file.  The \texttt{think-timeout} is
particularly important in our case because it directly corresponds to
the `impatience' of customers.  \texttt{httperf} (up to version 0.8)
supported only fixed values for these timeouts. We modified the code
to implement exponentially distributed \texttt{--think-timeout}
values.  This allowed us to use variable and random timeouts in our
experiments to enable us to more reasonably model user impatience.
Note that most other experimental works assume fixed timeout values.

Since we are modeling abandonments by timeouts, we must also model the
user behavior of retrying an abandoned request. The retry model that
we use is as follows. Whenever a request times out, it retries with a
probability of $p$ and abandons with probability of $1-p$.  The number
of retries per request is upper bounded by $M$. We added this new
functionality, which is accessed with the \texttt{--retry-model}
option, to \texttt{httperf}.

If any request in a session fails, even after the retries, the entire
session is considered to have failed. The remaining requests in that
session are not issued in such a case. This is the realistic model for
a Web-server because users would most likely `give up' and leave the
Web site, after failing to load a desired page. Thus, for a
transaction request to be generated, all the preceding browsing
requests of that session must have been completed successfully. This
clearly implies that to have a higher amount of revenue generation
under overload conditions, we must also increase the number of
browsing requests that are completed.  This would increase the chances
of success for a session that would result in a revenue-generating
transaction. Our proposal of giving a strictly higher priority to a
transaction request over browsing requests would then ensure that if a
transaction request is generated, it has a very high chance of
completion.

\begin{table}
    \centering
    \vspace*{0.1in}
    \begin {tabular}{|l|c|} \hline
      Request  & Mean execution time (mS) \newline \\ \hline
      Main Page (Br-1) & 200 \\
      Browsing (Br-2) & 300 \\
      Searching (Br-3) & 300 \\
      Details (Br-4) & 222 \\ \hline
      Login (Tr-1) & 280 \\
      Shipping (Tr-2) & 420 \\
      Payment (Tr-3) & 500 \\
      Confirm (Tr-4) & 300 \\ \hline
    \end{tabular}
    \caption{Mean execution time of CGI scripts.}
    \label{cgi-exe-time}
\end{table}

The server is configured with eight queues: four queues for browsing
requests and four for transaction requests. Thus each queue, and each
type of request has its own parameters and handling mechanisms. We
perform three sets of experiments as follows. 
\begin{enumerate}
\item \emph{SQ}: Single queue to store all the requests and served
  according to FIFO. The queue is assumed to have capacity to queue
  100 requests.
\item \emph{8Q-AF}: Eight queues, one for each type of request with
  all of them always serving in FIFO order. The browsing queues are
  assumed to have capacity to store 50 requests while the transaction
  queues have capacity for 25 requests. 
\item \emph{8Q-LIFO-Pri}: Eight queues, one for each type of request
  with LIFO at overload for browsing queues and FIFO for transaction
  queues and dynamic priority. Buffer capacities are as in 8Q-AF.
\end{enumerate}

The utility of the browsing queues and the transaction queues is
assigned in proportion to the probability that a request of that type
eventually results in a `confirm' (Tr-4) transaction.  For instance,
from Fig.~\ref{bro_buy_stages}, the probability that a `browse' page
(Br-2) will lead to a final `confirm' transaction Tr-4 is 0.022,
whereas the probability that a `details' (Br-4) page will lead to a
final `confirm' transaction is 0.073.  The utility for each type of
request for the parameters shown in Fig.~\ref{bro_buy_stages} is shown
in Table~\ref{table:utility_values}. Also, note that we assign the
utility to the queues in such a way that the transaction requests
always have higher priority than the browsing ones while their
relative priority changes dynamically.

\begin{table}
    \centering
    \vspace*{0.1in}
    \begin{tabular}{|l|c|} \hline
      Request queue & Utility \\ \hline
      Main Page (Br-1) & 27  \\ 
      Browsing (Br-2) & 22 \\
      Searching (Br-3) & 36 \\
      Details (Br-4) & 73 \\ \hline
      Login (Tr-1) & 3650 \\
      Shipping (Tr-2) & 4050 \\
      Payment (Tr-3) & 4500 \\
      Confirm (Tr-4) & 5000 \\ \hline
    \end{tabular}
    \caption{Utility values for queues.}
    \label{table:utility_values}
\end{table}


The \texttt{--timeout} value for these experiments is 8~seconds and
the \texttt{--think-timeout} is exponentially distributed with a mean
of 12~seconds. Thus the mean total timeout value for receiving a reply
is 20~seconds. If a request is timed out, the retry mechanism
described earlier comes into operation. The request is retried with a
probability of 0.4 ($p=0.4$) whenever a timeout occurs, up to a
maximum of 5 ($M=5$) retries per request. Also, the upper threshold of
the CPU utilization for the switch-over from FIFO to LIFO in the
browsing queues is 0.99 while the lower threshold value for the change
from LIFO to FIFO is 0.95.


\subsubsection {Experimental Results}

\begin{figure}
  \includegraphics[width=2.2in,angle=270]{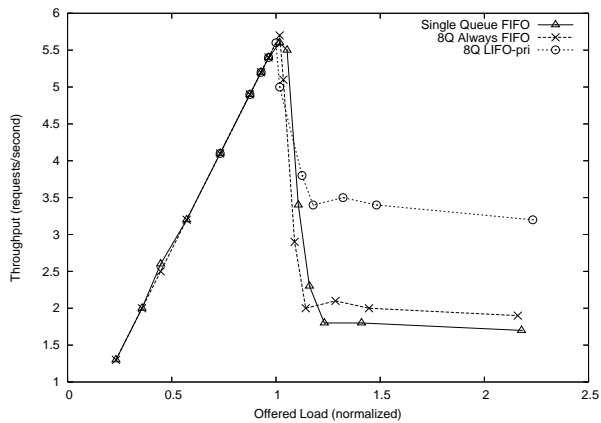}
  \caption{Overall throughput vs. load}
  \label{LOG_s_load_th}
\end{figure}

Fig.~\ref{LOG_s_load_th} shows the overall throughput as a function of
the offered load. We can see that when the load is below the capacity
of the server, i.e., $\rho=1$, (corresponds to 5.6 requests/second for
this workload model), all the three schemes have similar throughput.
When $\rho > 1$, the throughput of the {SQ} system drops significantly
and is the minimum of the three cases for $\rho > 1.3$. In the 8Q-AF
system a larger number of transaction requests complete and we can see
a marginal improvement in the throughput. The best performance is
clearly in the 8Q-LIFO-Pri system with a throughput of almost 3.5
requests/second (about 63\% of the server capacity) even for
$\rho=2.0$ .

\begin{table*}
  \begin{center}
    \vspace*{0.1in}
    \begin {tabular}{|c|l|l||r|r|r|r|r|} \hline 
      \textbf{$\rho$}& \textbf{Case}&\textbf{Requests} & \textbf{Browsing} & \textbf{Tr-1} &  \textbf{Tr-2} &  \textbf{Tr-3} & \textbf{ Tr-4} \\ \hline\hline 
      &\multicolumn{2}{|r||}{\textbf{Generated~~~}}  & 54480 & 888  &  792  &  720  & 648 \\\cline{2-8} \cline{2-8}
      & &\textbf{Completed }  & 54480 & 888  &  792  &  720  & 648  \\  \cline{4-8}
      &SQ& \textbf{Timed out} &   0  &  0   &  0   &  0   &  0 \\ \cline{4-8}
      & & \textbf{Dropped }  & \multicolumn{5}{|c|}{0} \\ \cline{2-8} \cline{2-8}
      & & \textbf{Completed}  & 54480 & 888  &  792  &  720  & 648 \\\cline{4-8}     
      0.85 & 8Q-AF& \textbf{Timed out} &  0  &  0   &  0   &  0   &  0 \\\cline{4-8}
      & & \textbf{Dropped }  &  0 &  0 & 0 & 0 & 0 \\ \cline{2-8} \cline{2-8}
      & &\textbf{Completed }  & 54480 & 888  &  792  &  720  & 648  \\  \cline{4-8}
      &8Q-LIFO-Pri& \textbf{Timed out} &  0  &  0   &  0   &  0   &  0 \\ \cline{4-8}
      & & \textbf{Dropped }  &  0 &  0 & 0 & 0 & 0 \\ \hline \hline
      & & \textbf{Generated}  & \multicolumn{5}{|c|}{42029} \\ \cline{4-8}
      & &\textbf{Completed}  & 16170  &    20  &   15  &  9  &  8 \\\cline{4-8}
      &SQ&\textbf{Timed out} & 20029 &  18  &   5  &  1  &  1 \\\cline{4-8}
      & & \textbf{Dropped }  &\multicolumn{5}{|c|}{5753}\\ \cline{2-8} \cline{2-8}
      & & \textbf{Generated} & 43324 & 24  & 20  &  19   &  15 \\  \cline{4-8}
      & & \textbf{Completed} & 19852 & 23  & 19  &  19   &  15 \\  \cline{4-8}
      1.4&8Q-AF& \textbf{ Timed out}   & 16305 & 1 & 1  & 0  & 0 \\\cline{4-8}
      & & \textbf{Dropped }  & 7167 & 0 & 0 & 0 & 0 \\ \cline{2-8} \cline{2-8}
      & & \textbf{Generated}  & 44826 &  195 & 137 & 99 & 53 \\ \cline{4-8}
      & & \textbf{Completed}  & 30851 &  187 & 127 & 87 & 50 \\ \cline{4-8}
      &8Q-LIFO-Pri&  \textbf{Timed out } & 4075  &  8  &  10   &  12 &  3 \\\cline{4-8}
      & & \textbf{Dropped }  & 9900 &  0 & 0 & 0 & 0 \\ \hline \hline
    \end{tabular}
    \caption{Throughput data for the different types of requests for
      different values of $\rho$. } 
    \label{LOG_s_results_all}
  \end{center}
\end{table*}

\begin{table*}
  \begin{center}
    \vspace*{0.1in}
    \begin {tabular}{|l|c|c|c||c|c|c|} \hline
      $\rho$ &  \multicolumn{3}{|c|}{$\rho=0.85$} & \multicolumn{3}{|c|}{$\rho=1.4$} \\ \hline
      Case& SQ & 8Q-AF & 8Q-LIFO-Pri & SQ & 8Q-AF & 8Q-LIFO-Pri \\ \hline
      Completed &100&100&100& 29.9 & 36.6 & 57.5 \\
      Timed out &  0&  0&  0& 36.8 & 29.9 & 7.5 \\
      Dropped   &  0&  0&  0& 10.6 & 13.1 & 18.2 \\
      Not Generated &  0&  0&  0& 22.8 & 20.4 & 16.8 \\ \hline
    \end{tabular}
    \caption{Throughput data (in percentage) for different values of $\rho$.}
    \label{LOG_s_rate}
  \end{center}
\end{table*}


We now discuss the results in more detail and analyze it at the
requests level. Table~\ref{LOG_s_results_all} shows the composition of
requests for each value $\rho$, along with the number of requests
completed, requests timed out, and requests dropped for each scheme
({SQ}, {8Q-AF} and {8Q-LIFO-Pri}) from each of the
queues. For 8Q-AF and 8Q-LIFO-Pri, the data for the browsing queues is
combined. Table~\ref{LOG_s_rate} shows the overall percentage of
requests completed, requests dropped, requests timed out and requests
that were not generated because the session aborted before completion.

We can see that when the offered load is less then the capacity of the
server, ($\rho=0.85$ case) the percentages are the same in all the
three schemes with 100\% of the sessions getting completed.

When the offered load exceeds server capacity, requests timeout and
generate retries which further increases the offered load to the
server. However, since some sessions are aborted, the requests after
the session abortion are not offered and this can cause some reduction
in the offered load. This effect is seen in the reduced number of
browsing and transaction requests \emph{generated} under each
policy---42,029 requests are generated in SQ as compared to 43,402 in
8Q-AF and 45,310 requests in 8Q-LIFO-Pri. The end result is that the
number of the Tr-4 requests (the \emph{direct} revenue-generating
request) completed\footnote{The seemingly disproportionate
decrease in these numbers as compared to the non-overload case can be
attributed to the lack of a load generator that could generate our
randomly distributed workload. However, the numbers are sufficient for
highlighting the performance improvement in LIFO-Pri as compared to
other schemes in overload conditions.} increases from 8 in SQ to 15 in
8Q-AF and to 50 in 8Q-LIFO-Pri. Recall that this number should be the
primary measure of performance of an E-commerce Web-server.

Our experimental setup represents the fact that browsing requests are
important in the sense that they are the \emph{source} of transaction
arrivals. Increasing the browsing request completion rate, coupled
with priority to transaction service, results in an overall increase
in the transaction completion rate.

Table~\ref{LOG_s_results_all} shows that with $\rho=1.4$, the
{LIFO-Pri} scheme increases the number of `login' requests
generated to 195, out of which 187 are actually completed, only 8 time
out and there are zero drops.  This is due to the fact that a larger
number of browsing requests are completed, which in turn leads to the
generation of transaction requests.

\begin{figure}
    \centering
    \includegraphics[width=2.2in,angle=270]{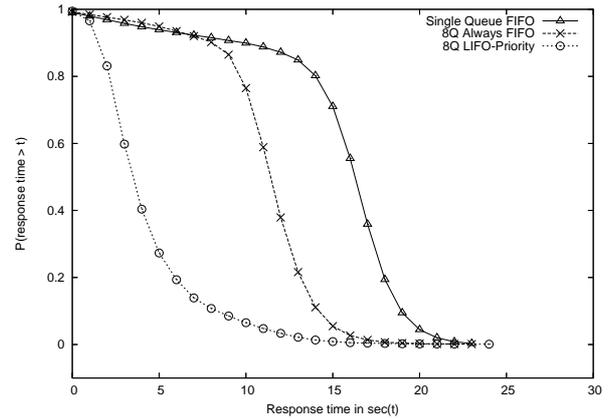}
    \caption{Response time distribution for `main' page (Br-1) for $\rho=1.4$}
    \label{resp-dist-br1-high}
\end{figure}

Some more observations from Table~\ref{LOG_s_results_all}:
\begin{itemize}
\item The number of timed-out transactions is higher for LIFO-Pri than
  for 8Q-AF. Although this may seem surprising, observe that that a
  significantly larger number were generated, e.g., 195 Tr-1 requests
  for LIFO-Pri as compared to 24 for 8Q-AF. 
\item The effect of LIFO on reducing abandonments is clearer from the
  browsing requests where the difference in the number generated is
  not very significant (44,826 \textit{vs.} 43,324). However, only
  19,852 completed in 8Q-AF \textit{vs.} 30,851 in LIFO-Pri---a result
  of the reduction of the number of request abandonments from 16,305
  to 4,075.
\item For 8Q-AF no transaction requests are dropped even at high loads
  because these queues have a high priority and also because very few
  are offered.
\item Using LIFO in the browsing queues along with priority for
  transaction queues as in 8Q-LIFO-Pri retains the benefits of giving
  high priority to the transaction requests. This can be seen in
  Table~\ref{LOG_s_results_all} where the number of transaction
  requests dropped in 8Q-LIFO-Pri is zero even in overload conditions.
\end{itemize}

\begin{figure} 
    \centering
    \includegraphics[width=2.2in,angle=270]{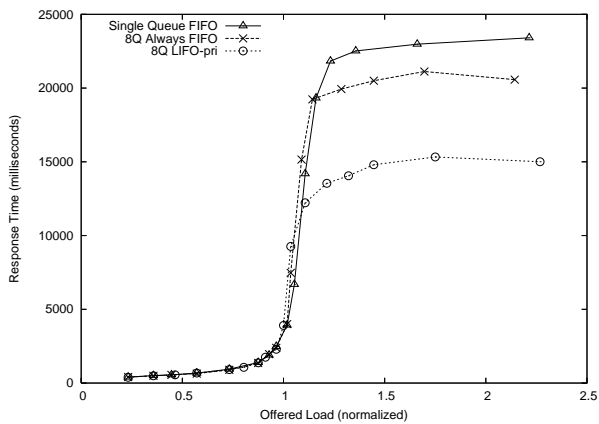} 
    \caption{Average response time vs. load}
    \label{LOG_s_load_resp}
\end{figure}

Fig.~\ref{resp-dist-br1-high} shows the response time distribution of
Br-1 requests for $\rho=1.4$; we see that for 8Q-LIFO-Pri, nearly 80\%
of the requests have a response time less then 5~seconds, whereas in
SQ and in 8Q-AF only about 10\% of the requests achieve this.

Fig.~\ref{LOG_s_load_resp} shows the graph between the average
response time (of completed transactions) as a function of $\rho$ for
the three policies. The response time with the LIFO-Pri policy is
significantly better during overload. Given the improved throughput
performance of LIFO-Pri, as was observed from
Table~\ref{LOG_s_results_all}, this is not surprising because in the
presence of abandonments it is necessary to improve response time
performance to be able to increase throughput. Improving response time
reduces request abandonments, which in turn causes fewer session
abandonments and an increased overall throughput.

\balance 
\section{Summary and Discussion}
\label{sec:Conclude}

In this paper, we proposed and experimentally evaluated an
overload control scheme for Web-servers under a reasonably realistic
model of for E-commerce workload. The LIFO-Pri scheme proposed in this
paper is an extremely simple, yet effective, mechanism for overload
control.  The experimental results are highly encouraging---the server
could operate at nearly 60\% of its maximum capacity even when offered
a load 1.5 times its capacity and has a factor of 7 increase in the
number of direct revenue-generating requests completed as compared to
a single queue model during overload.

The benefits of LIFO were observed by Dalal and Jordan~\cite{Dalal01},
however, the results were not for an E-commerce environment, and no
implementation and experiments were done (validation was by
simulation). We believe our work confirms experimentally the truly
remarkable effect on performance during overload of the LIFO policy
along with a priority for revenue-generating requests. Although the
LIFO service policy seems to always imply high variability and
unfairness, the abandonment and retry behavior of users during
overload, turns LIFO into a compelling choice.

Future work includes having better indicators for overload---this work
assumed that the CPU was the bottleneck resource and used CPU
utilization as the indicator. We would like to extend this to cases
where we do not know the bottleneck resource. Work is also needed to
model user behavior even more appropriately (e.g. longer response
times should discourage `repeat' visits). Lastly, analytical models
are necessary to gain further insight into overload control mechanisms
for Web-servers.

\bibliographystyle{ieee}
\bibliography{main}


\end{document}